\documentclass[aps,prl,twocolumn,superscriptaddress,showpacs,10pt]{revtex4-1}
\usepackage{graphicx}
\usepackage{amssymb}
\usepackage{amsmath}
\pdfoutput=1

\begin{document}

\def\pc{\frac{2\pi}{\Phi_0}}

\def\e{\varepsilon}
\def\f{\varphi}
\def\p{\partial}
\def\ba{\mathbf{a}}
\def\bA{\mathbf{A}}
\def\bb{\mathbf{b}}
\def\bB{\mathbf{B}}
\def\bD{\mathbf{D}}
\def\be{\mathbf{e}}
\def\bE{\mathbf{E}}
\def\bH{\mathbf{H}}
\def\bj{\mathbf{j}}
\def\bk{\mathbf{k}}
\def\bK{\mathbf{K}}
\def\bM{\mathbf{M}}
\def\bm{\mathbf{m}}
\def\bn{\mathbf{n}}
\def\bq{\mathbf{q}}
\def\bp{\mathbf{p}}
\def\bP{\mathbf{P}}
\def\br{\mathbf{r}}
\def\bR{\mathbf{R}}
\def\bS{\mathbf{S}}
\def\bu{\mathbf{u}}
\def\bv{\mathbf{v}}
\def\bV{\mathbf{V}}
\def\bw{\mathbf{w}}
\def\bx{\mathbf{x}}
\def\by{\mathbf{y}}
\def\bz{\mathbf{z}}
\def\Bn{\boldsymbol{\nabla}}
\def\Bo{\boldsymbol{\omega}}
\def\Br{\boldsymbol{\rho}}
\def\Bs{\boldsymbol{\hat{\sigma}}}
\def\bh{{\beta\hbar}}
\def\mA{\mathcal{A}}
\def\mB{\mathcal{B}}
\def\mD{\mathcal{D}}
\def\mF{\mathcal{F}}
\def\mG{\mathcal{G}}
\def\mH{\mathcal{H}}
\def\mI{\mathcal{I}}
\def\mL{\mathcal{L}}
\def\mO{\mathcal{O}}
\def\mP{\mathcal{P}}
\def\mS{\mathcal{S}}
\def\mT{\mathcal{T}}
\def\mU{\mathcal{U}}
\def\mV{\mathcal{V}}
\def\mZ{\mathcal{Z}}
\def\fr{\mathfrak{r}}
\def\ft{\mathfrak{t}}
\def\tV{\tilde{V}}
\def\tN{\tilde{N}}
\newcommand{\rf}[1]{(\ref{#1})}
\newcommand{\al}[1]{\begin{aligned}#1\end{aligned}}
\newcommand{\ar}[2]{\begin{array}{#1}#2\end{array}}
\newcommand{\eq}[1]{\begin{equation}#1\end{equation}}
\newcommand{\bra}[1]{\langle{#1}|}
\newcommand{\ket}[1]{|{#1}\rangle}
\newcommand{\av}[1]{\langle{#1}\rangle}
\newcommand{\AV}[1]{\left\langle{#1}\right\rangle}
\newcommand{\braket}[2]{\langle{#1}|{#2}\rangle}
\newcommand{\ff}[4]{\parbox{#1mm}{\begin{center}\begin{fmfgraph*}(#2,#3)#4\end{fmfgraph*}\end{center}}}

\newcommand{\degree}{\ensuremath{^\circ}}
\newcommand{\comment}[1]{}

\def\mr{m_{\perp}}
\def\ml{m_{\parallel}}
\def\hr{H_{\perp}}
\def\hl{H_{\parallel}}

\def\mb{(\mu+\alpha\nu)}
\def\nb{(\nu-\alpha\mu)}
\def\lb{(\lambda+\alpha\kappa)}
\def\kb{(\kappa-\alpha\lambda)}
\def\mn{\left|\bm\times\bz\right|}
\def\etap{\frac{2\pi}{\Phi_0}}
\def\ab{\bar{\alpha}}

\title{Spin-torque ac impedance in magnetic tunnel junctions}

\author{Silas Hoffman}
\affiliation{Department of Physics and Astronomy, University of California, Los Angeles, California 90095, USA}

\author{Pramey Upadhyaya}
\affiliation{Department of Electrical Engineering, University of California, Los Angeles, California 90095, USA }

\author{Yaroslav Tserkovnyak}
\affiliation{Department of Physics and Astronomy, University of California, Los Angeles, California 90095, USA}

\begin{abstract}

Subjecting a magnetic tunnel junction (MTJ) to a spin current and/or electric voltage induces magnetic precession, which can reciprocally pump current through the circuit. This results in an ac impedance, which is sensitive to the magnetic field applied to the MTJ. Measuring this impedance can be used to characterize the coupling between the magnetic free layer and the electric current as well as a read-out of the magnetic configuration of the MTJ.

\end{abstract}

\maketitle
The development of the next generation of computer memory and logic can be made possible by current-driven effects through magnetic multilayers by utilizing the mechanisms of tunnel magnetoresistance \cite{jullierePLA75} and spin-transfer torque \cite{slonczewskiJMMM96,*bergerPRB96}. These effects have been demonstrated to efficiently read and write bits as furnished by magnetic domains \cite{ikedaIEEE07,*chenIEEE10}. Somewhat less utilized are the recently discovered torques due to the voltage-induced anisotropy \cite{weisheitSCI07,*maruyamaNATNANO09}. The applied voltage induces a charge build up at the tunnel-barrier interface with the transition-metal ferromagnet. The strong electric field at the interface modifies electronic structure along with the local occupation of the $d$-character bands in the transition metal. Due to spin-orbit coupling, this results in the anisotropic interaction between the local excess charge and the magnetization. Conventional MTJ spin-torque devices can benefit from the voltage-induced anisotropy by reducing the critical switching current for a fixed thermal stability \cite{wangNATMAT11}. Recently, it has been shown that this voltage-controlled magnetic anisotropy (VCMA) can induce ferromagnetic resonance \cite{nozakiNATPHYS12,zhuPRL12} or reverse the magnetic direction \cite{shiotaNATMAT11}. 

\begin{figure}[pt]
\includegraphics[width=\linewidth,clip=]{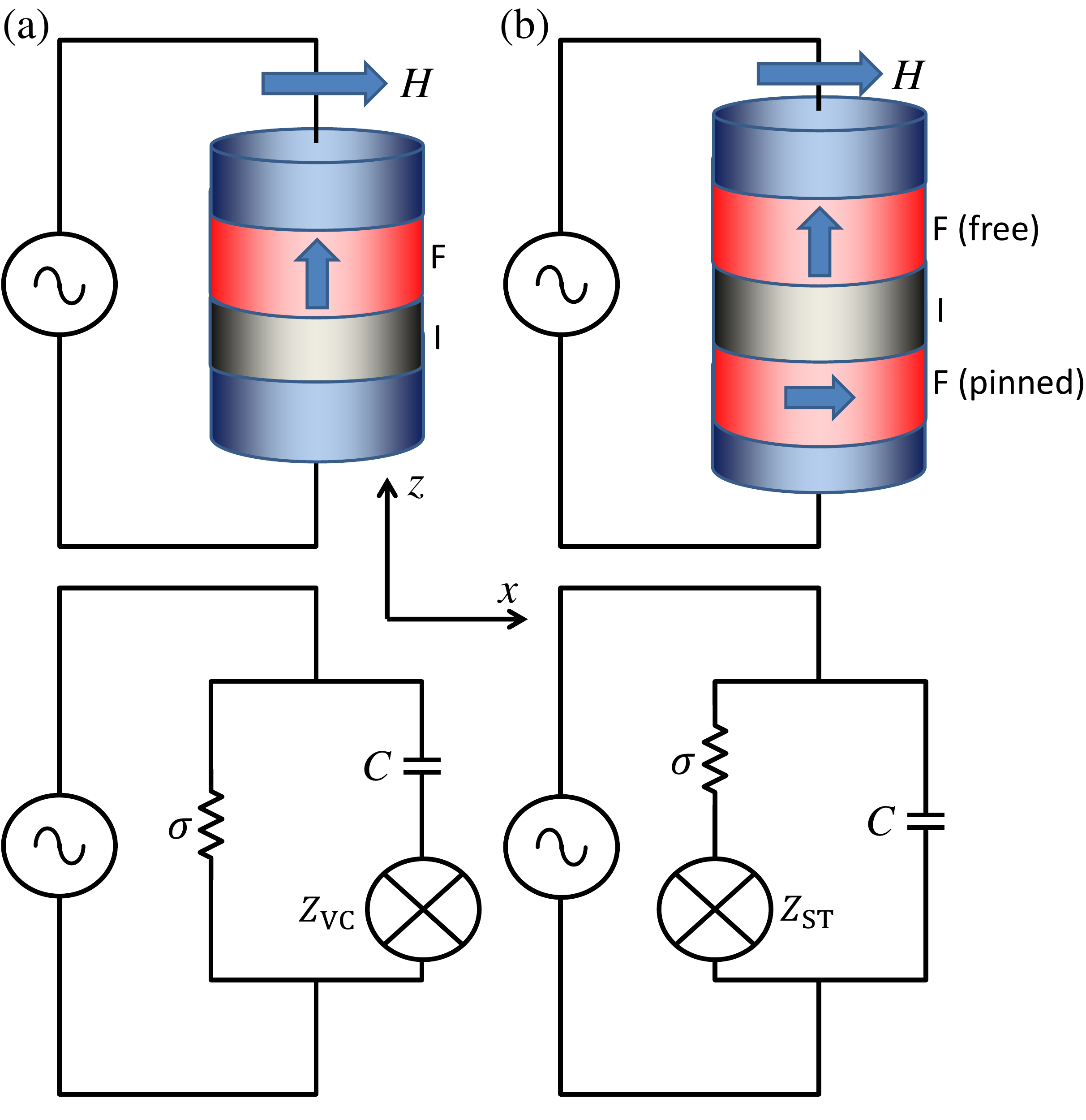}
\caption{Schematics of magnetic tunnel junction subjected to a magnetic field, as part of an ac  circuit that drives magnetic precession by VCMA (a) or Slonczewski torque (b), and the equivalent circuit diagrams (below) showing the additional impedance due to pumping by magnetic dynamics.}
\label{fig1}
\end{figure}

Here, we include the reciprocal backaction of magnetic dynamics on the circuit. Applying an ac voltage drives precession of the magnet that in turn pumps current, contributing to the ac impedance. In order to illustrate two physically distinct mechanisms of voltage-induced spin torques, we consider two special MTJ structures: (a) a ferromagnet (F)$\mid$insulator (I) bilayer with interfacial spin-orbit interaction and (b) an F$\mid$I$\mid$F heterostructure where one of the ferromagnetic layers is pinned and the other free. See Fig.~\ref{fig1} for schematics. In Fig.~\ref{fig1}(a), the ferromagnet is agitated by voltage-induced anisotropy \cite{weisheitSCI07,*maruyamaNATNANO09}, while in Fig.~\ref{fig1}(b) the free magnetic layer is driven by spin-transfer torque \cite{slonczewskiJMMM96,*bergerPRB96}. In practice, these two scenarios can be accessed by varying the thickness of the spacer: for thicker spacers that are Ohmically opaque the VCMA must ultimately dominate, while for thinner spacers the spin-transfer torque should become progressively more important. In the former case, within our model, the impedance vanishes when the equilibrium magnetization is parallel or perpendicular to the direction of induced anisotropy. In the latter case, we find that the impedance is enhanced when the magnetic equilibrium is nearly perpendicular to the direction of polarization of the spin current. If the magnitude of the VCMA and spin-transfer torque are comparable, we can tune between these effects by applying a magnetic field. The resultant impedance shift can be used to characterize the magnitude and the nature of the coupling between ferromagnet and circuit, as well as to probe magnetic configuration.

The following analysis of voltage-controlled magnetic anisotropies [Fig.~\ref{fig1}(a)] applies to a general class of MTJ's that break mirror symmetry normal to the face of the magnetic layer, inducing Rashba-type spin-orbit interaction \cite{bychkovJCP84}. Consider an F$\mid$I bilayer subjected to a voltage $V$ in the external circuit, wherein the free energy of the transferred charge $Q$ is $F[Q]=-QV$. We treat the ferromagnetic layer to be monodomain with free-energy density
\eq{F[\bM]/\mathcal{V}=\frac{1}{2}(N_x M_x^2+N_y M_y^2+N_z M_z^2 - K M_z^2)- H M_x\,,}
where $\mathcal{V}$ is the volume of the ferromagnetic layer and $\mathbf{M}=(M_x,M_y,M_z)$ is the magnetization vector. To be specific, we take the $xy$ cross section of the magnet to be an ellipse elongated in the $x$ direction. $N_x+N_y+N_z=4\pi$ are the demagnetization factors (with $N_y>N_x$), $H$ is the applied field along the semimajor ($x$) axis, and $K$ is the perpendicular anisotropy that can be induced by the insulating layer in the absence of voltage. We consider a geometry wherein the perpendicular anisotropy overcomes the long-range dipole field such that $N_x>N_z-K$ and, when no magnetic field or voltage is applied, the equilibrium orientation of the magnet is perpendicular to the interface ($z$ axis). Under application of a magnetic field along the semimajor axis of the ellipse, the equilibrium magnetization tilts away from the $z$ axis as $\bar{\bM}=(H /\tilde{N}_x,0,\sqrt{M_s^2-(H /\tilde{N}_x)^2})$, where $M_s=|\bM|$ is the saturated magnetization, $\tN_x=N_x-N_z+K$ and $\tN_y=N_y-N_z+K$. Note that $\tN_x,\tN_y>0$ are guaranteed by $N_y>N_x>N_z-K$. For simplicity, we restrict $\left|H\right|<M_s\tilde{N}_x$.

The tunneling layer is treated as a parallel plate capacitor of capacitance $C$, storing energy $F[q]=q^2/2C$ where $q$ is the charge on the surface of the insulator. The structure of the device breaks mirror symmetry in the direction perpendicular to the interface. We treat the lateral dimensions of the device macroscopically, as compared with the microscopic spin-orbit interactions inducing VCMA, and thus require the coupling to be rotationally symmetric around the $z$ axis. Subject to these symmetries, the anisotropy controlled by voltage must be induced in the direction of the broken mirror symmetry. To satisfy time reversal symmetry, the free energy must be of even order in magnetization. Since there are no Ohmic losses associated with tunneling through the junction, we suppose the dominant interaction between the magnetization and the electric circuit to be nondissipative. Dissipative corrections could be taken into account similarly to Ref.~\cite{hoffmanPRB12}. Because this torque is induced by the electric field at the interface, we take the energy to be proportional to electric flux in the insulating layer. The coupling, up to quadratic order in magnetization, is $F[\bM,q]=-\nu q m_z^2/2$ where $\nu$ is the phenomenological coupling between the projection of the magnetic direction along the $z$ axis, $m_z=M_z/M_s$, and the circuit. The full free energy is the sum of these individual components $F=F[\bM]+F[Q]+F[q]+F[\bM,q]$.

The equation of motion of the ferromagnet is described by the Landau-Lifshitz-Gilbert (LLG) equation \cite{landauBK80,*gilbertIEEEM04}
\eq{\dot{\bm}=-\gamma\bm\times\bH+\alpha\bm\times\dot{\bm}\,,}
where $\gamma$ is the gyromagnetic ratio and $\alpha$ is the dimensionless Gilbert damping. The magnetic direction vector is $\bm=\textbf{M}/M_s$ and $\bH=-\mathcal{V}^{-1}\partial F/\partial \textbf{M}$ is the effective field. Applying a small ac voltage to the circuit induces precession of the magnet due to the VCMA. Treating the response to voltage linearly, the solution to the equations of motion for the magnet is a damped harmonic oscillator, driven by voltage, centered around the equilibrium value of magnetization,  $\bar\bm=(\bar m_x,\bar m_y,\bar m_z)$, with resonance $\omega_0=\gamma M_s\sqrt{\bar m_z^2\tN_x\tN_y}$. To obtain an expression of the current through the circuit $\dot Q$, we note that the difference in change in charge between the reservoir and the capacitor is the leakage current due to tunneling through the insulator $\dot Q - \dot q=\sigma V$ where $\sigma$ is the junction conductance (disregarding magnetoconductance). Neglecting impedance in the external circuit, $\partial_Q F = -\partial_q F$, we find
\eq{\dot Q=C\dot V+\sigma V + \nu C m_z\dot m_z\,.}
In addition to the resistor and capacitor in parallel, the precession of the magnet pumps current, which is reciprocal to the VCMA. For the resonant driving at $\omega_0$, the additional impedance, informed by the circuit diagram in Fig.~\ref{fig1}(a), is
\eq{Z_{\rm{VC}} = \frac{\nu^2}{\alpha\omega_0^2 S} \frac{\bar m_x^2 \bar m_z^2\tN_y}{\tN_y+\bar m_z^2\tN_x}\,,
\label{Zvc}}
where $S=\mV M_s/\gamma$ is the total spin angular momentum. We have suppressed higher-order terms in $\alpha$, assuming $\alpha\ll1$, which it typically the case in practice. Notice that on resonance $Z_{\rm{VC}}$ is real, but away from resonance it is generally complex valued. The impedance is second order in $\nu$ reflecting the VCMA driving of the magnet and subsequent self-consistent pumping by magnetic precession. $Z_{\rm VC}$ is proportional to the product of the equilibrium value of magnetization along the $x$ and $z$ axes, being maximized at an intermediate polar angle, and can therefore be modulated by applied magnetic field. The effect is larger for smaller Gilbert damping.

Let us now consider an MTJ wherein the torque is induced by spin current polarized along the direction of the pinned layer ($x$ axis) \cite{slonczewskiJMMM96,*bergerPRB96}, as sketched in Fig.~\ref{fig1}(b). (Note that one can obtain a similar effect also in an F$\mid$I bilayer due to spin-orbit interaction \cite{manchonPRB08,*chernyshovNAT09,*halsEPL10}.) The equation of motion of the magnet coupled to an external circuit by Slonczewski torque is
\eq{
\dot\bm=-\gamma\bm\times\bH+\alpha\bm\times\dot\bm+\mu(\dot Q-\dot q)\bm\times\bx\times\bm\,,
\label{llgstt}
}
where $\mu=S(\hbar/2 e)P/(1+\bar m_xP^2)$ \cite{slonczewskiPRB05}, as determined by microscopic considerations, characterizes the strength of the torque induced by current on the magnet. $P$ is the tunneling spin polarization. The torque is proportional to the leakage current, $\dot Q-\dot q$, through the capacitor. The equation of motion for charge satisfying microscopic time-reversal symmetry consistent with Eq.~(\ref{llgstt}) is \cite{tserkovPRB08tb}
\eq{\dot Q=\sigma (V-\mu S\dot\bm\cdot\bm\times\bx)+C\dot V\,.}
In contrast to the above VCMA model, $\sigma=\sigma_0(1+\bar{m}_xP^2)$, the tunnel magnetoconductance, depends on the relative
orientation of the pinned and free magnetic layers. Similar to the previous case, we apply an ac voltage at resonance which drives magnetic dynamics and shifts the impedance due to the charge pumping from magnetic precession. Again neglecting terms higher-order in $\alpha$, the associated impedance is
\eq{Z_{\rm{ST}}=-\frac{\mu^2 S}{\alpha} \frac{\bar m_z^4\tN_x}{\tN_y+\bar m_z^2\tN_x}\,.
\label{Zst}}
Although at resonance $Z_{\rm{ST}}$ is real and negative, the second law of thermodynamics bounds the Slonczewski torque parameter $\mu^2\leq\alpha/\sigma S$ \cite{hoffmanPRB12}, ensuring positivity of $1/\sigma+Z_{\rm{ST}}$. Note that $ Z_{\rm{ST}}$ vanishes when the equilibrium magnetization is perpendicular to the $z$ axis. 

\begin{figure}[pt]
\includegraphics[width=\linewidth,clip=]{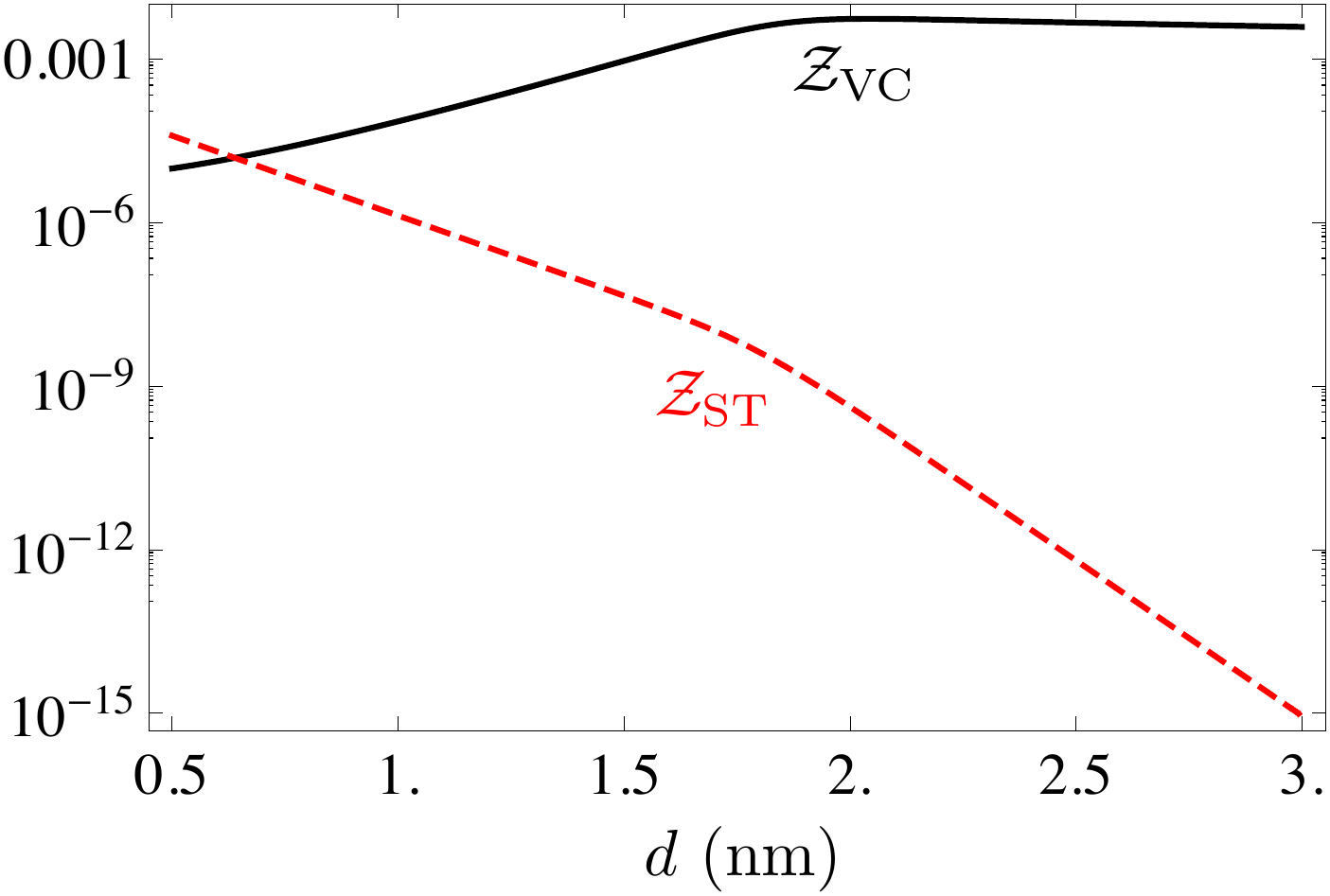}
\caption{Relative change in impedance as a result of VCMA (black solid) or Slonczewski torque (red dashed) as a function of  MgO spacer thickness, evaluated at $\bar m_x=0.73$ and $\bar m_x=0$, respectively. The kinks near $d\approx2$~nm correspond to the crossover at $\sigma\sim\omega_0 C$.}
\label{fig2}
\end{figure}

We now make an estimate of the effect for practical memory devices. Specifically, we consider a $150\times70$~nm$^2$ elliptical nanopillar of CoFeB/MgO, where the thickness of the magnetic layer is $1.6$~nm, giving $N_x=0.2$ and $N_y=0.5$. The other relevant parameters are taken from Ref.~\cite{zhuPRL12}: $\alpha=3\times10^{-2}$, $M_s=950$~emu/cm$^3$, $\nu=2.8$~$\mu$erg statV$^{-1}$ cm$^{-1}$, $K=12$, $P=0.5$, and $\sigma_0=1.6$~mS at the MgO thickness of $d=0.86$~nm (with the exponential decay length as a function of $d$ of $0.15$~nm \cite{amiriAPL11}). Our figure of merit is the relative change in impedance as a result of  VCMA or spin-transfer torque: $\mZ\equiv |Z-Z_0|/|Z_0|$, where $Z$ is the total impedance of the circuit in the presence of the magnetic dynamics and $Z_0$ is the impedance of the static junction (taking dielectric constant of MgO to be $\epsilon=10$). To evaluate $\mZ_{\rm VC}$ and $\mZ_{\rm ST}$, we first choose the magnetic field $H$ such that $\bar m_x$ maximizes the dimensionless geometric factors in Eq.~\eqref{Zvc} and Eq.~\eqref{Zst}: $\bar m_x=0.73$ and $\bar m_x=0$, respectively. Varying the tunnel barrier thickness $d$ at a fixed magnetic field, we plot the corresponding $\mZ$ in Fig.~\ref{fig2}. $\mZ_{\rm{VC}}$ increases exponentially reaching the maximum at $\sim 1\%$ near $d\approx 2$~nm. Past this thickness, the junction ac behavior crosses over from the resistive to capacitive regime, in which $\mZ_{\rm{VC}}$ falls off inversely with $d$. Since spin-transfer torque is roughly proportional to conductance for thin barriers, $\mZ_{\rm{ST}}\propto\sigma$ decreases exponentially with increasing spacer thickness, for $d\lesssim2$~nm. For thicker barriers, $\mZ_{\rm{ST}}$ becomes proportional to $\sigma^2$, doubling the logarithmic slope in its $d$ dependence. For the smallest feasible spacer thickness of $0.5$~nm, the relative change in impedance is $\sim10^{-4}$.

\begin{figure}[pt]
\includegraphics[width=0.97\linewidth,clip=]{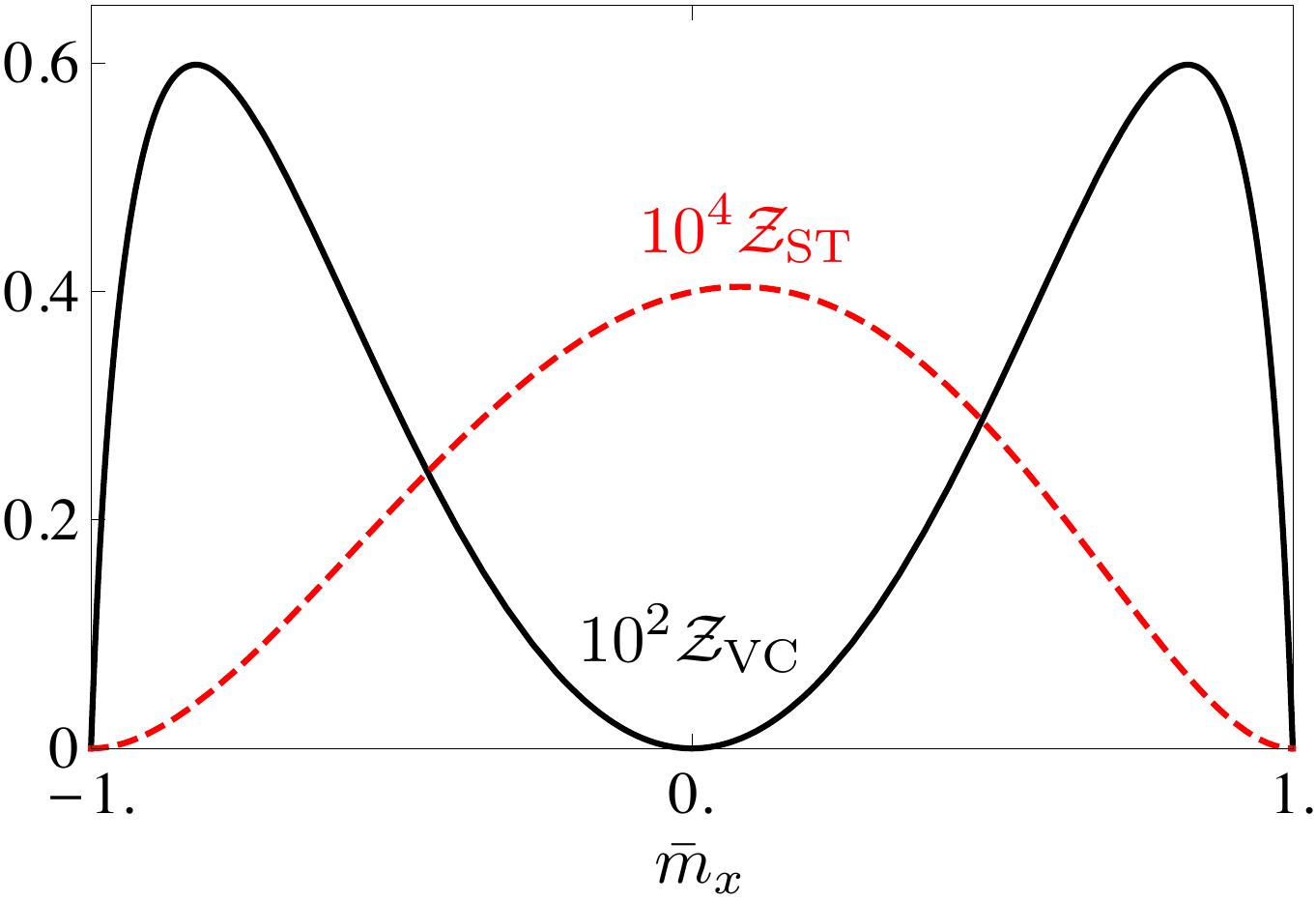}
\caption{Relative change in impedance as a result of VCMA (black solid) or Slonczewski torque (red dashed) as a function of $\bar{m}_x$ at the MgO spacer thicknesses $d=2$~nm and $d=0.5$~nm, respectively. The voltage is assumed to be applied at the frequency of ferromagnetic resonance, $\omega_0$, which depends on $\bar{m}_x$.}
\label{fig3}
\end{figure}

Next, allowing the external magnetic field to vary, we plot $\mZ_{\rm VC}$ and $\mZ_{\rm ST}$ as a function of equilibrium value of magnetization in Fig.~\ref{fig3}, fixing $d=2$~nm and $d=0.5$~nm, respectively. (These choices for $d$ are motivated by the respective maxima in $\mZ$ in Fig.~\ref{fig2}.) In the case of Slonczewski torque, there is a small asymmetry in the function $\mZ(\bar{m}_x)$ around zero due to variation of $\sigma$ and $\mu$ with $\bar{m}_x$. Note that the functional dependences of $\mZ_{\rm VC}$ and $\mZ_{\rm ST}$ on $\bar{m}_x$ are qualitatively distinct (with the former having double-lobe and the latter single-lobe profiles), allowing for a clear experimental differentiation between VCMA and spin-transfer torque regimes by measuring ac impedance as a function of magnetic field.

Measuring the ac impedance shift due to resonant magnetic dynamics could be an efficient method for characterizing the magnitude of the coupling between voltage and ferromagnet ($\nu$) and current and ferromagnet ($\mu$), as well as distinguishing between the two scenarios. Furthermore, when the thickness of the spacer is large, and thus the conductance is prohibitively small to utilize tunnel magnetoresistance, one could envision that measuring the ac impedance may be used as a nondestructive low-dissipation bit read-out. Owing to the promising energy efficiency of VCMA and the reciprocal effect, we expect an active search for ferromagnet-insulator interfaces with higher values of $\nu$, which could electrically control and read the direction of the magnet without tunneling current.

The authors would like to thank Juan G. Alzate, Pedram Khalili Amiri, and Scott Bender for their stimulating input. This work was supported in part by the DARPA and the NSF under Grant No. DMR-0840965.

\end{document}